\documentclass[12pt,english,floatfix,superscriptaddress,aps,prd,preprint]{revtex4}
\usepackage[utf8]{inputenc}
\usepackage{amsmath}
\usepackage{amssymb}
\usepackage{amsbsy}
\usepackage{amsfonts}
\usepackage{amsopn}
\usepackage{amstext}
\usepackage{graphicx}
\usepackage{amssymb}
\usepackage{amsfonts}
\usepackage{amsmath}
\usepackage{amsmath,amsthm,amsfonts,amssymb}
\usepackage[mathcal]{eucal}
\usepackage{mathrsfs}
\usepackage{graphicx}
\usepackage[english]{babel}
\usepackage{color}
\usepackage{esint}
\usepackage[dvips]{epsfig}
\usepackage[dvips]{graphicx}
\usepackage{float}
\usepackage{units}
\usepackage{textcomp}

\usepackage{hyperref}
\usepackage{slashed}

\newcommand{\ie}{\begin{equation}}
\newcommand{\fe}{\end{equation}}
\newcommand{\se}{\begin{eqnarray}}
\newcommand{\ff}{\end{eqnarray}}

\usepackage{xcolor}
\usepackage{soul}

\begin{document}

\title{Continuously transforming kinks into compactons in the $O(3)$-sigma model}


\author{F. C. E. Lima}
\email{cleiton.estevao@fisica.ufc.br}
\affiliation{Universidade Federal do Cear\'a (UFC), Departamento de F\'isica,\\ Campus do Pici, Fortaleza - CE, C.P. 6030, 60455-760 - Brazil.}


\author{D. A. Gomes}
\email{deboragomes@fisica.ufc.br}
\affiliation{Universidade Federal do Cear\'a (UFC), Departamento de F\'isica,\\ Campus do Pici, Fortaleza - CE, C.P. 6030, 60455-760 - Brazil.}


\author{C. A. S. Almeida}
\email{carlos@fisica.ufc.br}
\affiliation{Universidade Federal do Cear\'a (UFC), Departamento de F\'isica,\\ Campus do Pici, Fortaleza - CE, C.P. 6030, 60455-760 - Brazil.}

\date{\today}

\begin{abstract}
In this work, we investigate the solutions of vortices in the  $O(3)$-sigma model with the gauge field governed by the Chern-Simons term and subject to a hyperbolic self-dual potential. We show that this model admits both topological and nontopological solitons solutions. By means of numerical analysis, we realize that the topological solutions of the model can be transformed into compacton-like solutions. On the other hand, after modifying the model by the introduction of a dielectric constant, an interesting feature appears; namely,  the nontopological solutions can be transformed into kink-like solutions through the numerical variation of the dielectric constant. Finally, we discuss the degeneracy for the topological solitons in a given sector and present the numerical solutions of the first model.\end{abstract}

\maketitle

\section{Introduction}

The study of localized structures in $(2+1)D$ systems is relevant to the description of supercondutivity phenomena, since they essentially take place in planes \cite{1, Laughlin}. The investigation of the Chern-Simons term in these systems has become more common due the theoretical description of vortice structures \cite{Gosh}. Originally written in $(2+1)D$, the Chern-Simons term describes the eletromagnetic field dynamics with the usual Maxwell term \cite{Sales}. However, is worth noticing that the Chern-Simons term, when compared to the Maxwell term, is dominant in regions distant from eletromagnetic field sources \cite{Gosh, Sales}.

Theories with the Chern-Simons term have been vastly studied in the last decades \cite{horvathy,Cunha}. One remarkable feature of these theories is the fact that they have solutions with rotational symmetry. Such solutions represent localized and charged tubes of magnetic flux and in this way they correspond to the description of ideal anyons, which are objects obeying fractionary statistics \cite{1, Laughlin}. The Chern-Simons term is then believed to be linked to the fractionary statistics of these particles, suggesting the existence of a connection between the Chern-Simons term and the supercondutivity phenomenon at high temperatures \cite{Haldane, Zhang}.

An essential aspect of the study of two-dimensional models in $(2+1)D$ field theory is the existence of solutions of the solitary-like wave, that means, solitonic solutions \cite{Gosh, Casana1}. Solitons in two or multiple dimensions are extensively studied and have received increasing attention, not only for fundamental reasons but also for their applicability \cite{Leblond, Christ, Eisenberg}. For instance, we notice that, in $(1+1)D$, the soliton theory relies on the theory of completely integrable equations using the method of inverse dispersion transformation. The most common and relevant of these equations is a non-linear equation, the Schrödinger equation. In higher dimensions, for example, $(2+1)D$, the completely integrable equations are quite scarce \cite{Leblond, Morandotti, Fleischer}.

The interest in the study of $O(3)$-sigma models is due to its contribution to the description of the Heisenberg ferromagnetism \cite{Belavin}. The sigma model $O(3)$ in $(2+1)D$ is exactly integrable in the Bogomol'nyi limit \cite{Gosh} and the stability of these solitonic solutions is guaranteed by topological arguments \cite{Gosh, Cunha}. However, the solitons in this model can be expressed in terms of functions that are not scale invariant. As a consequence, the size of these solitons can change arbitrarily while  they evolve in time without modifying their energy \cite{Gosh, Samoilenka, CLee}.

Essentially, there are various ways of breaking the scale invariance in this model \cite{Leese, KLee}; the construction of {\it Q-lumps} is an example where the scale invariance is broken by including a specific potential term in the sigma model \cite{KLee}. Therefore, the size colapse of the soliton in this model is prevented by making a rotation in the inner space of the field variables. The above-mentioned solitons have finite energy and are consequently time dependent with a constant angular velocity \cite{Gosh}.

Recently, the study of the so-called compacton-like solutions has notably instigated the interest of some researchers \cite{Casana, Bazeia3, Bazeia4}. Compactons are finite wavelength solitons \cite{Rosenau} and they have been the subject of numerous studies since models containing topological defects can be employed to describe particles and cosmological objects, type cosmic strings \cite{Nielsen}. Another motivation for the expanding interest in investigating compact structures is that the compact vortices and skyrmions are intrinsically connected to spintronics \cite{Jubert, Romming}. Additionally, applications of the study of compact structures appear in braneworlds scenarios, and its properties are well discussed by Veras {\it et. al.} in Ref.  \cite{Veras}.

Theoretical models subject to hyperbolic potentials have emerged some years ago in some areas of physics, such as in the study of position-dependent mass systems in quantum mechanics as an attempt to describe the dynamics of solid state systems like abrupt heterojuctions and quantum dots \cite{CA1,CA2, Bastard}. Not very far from the quantum theories, the hyperbolic potential have also been used in the investigation of vortices solutions of the scalar field \cite{Bazeia1}. Furthermore, scalar fields subject to hyperbolic interactions can be applied to scalar and solvable black holes, for example, to find a regular configuration of non-charged black holes and the cosmological scalar field \cite{Bazeia2}. As discussed by Bazeia {\it et. al.} in Ref. \cite{Bazeia1, Bazeia2}, this field falsifies the Wheeler conjecture \cite{Ruffini} and avoids the scalar ``no-hair" theorem \cite{Bizon}. As a consequence, classes of exact solutions appear, including new scalar black holes with hyperbolic potentials \cite{Bazeia2}.

In this work, we discuss the existence of solitons in a minimal coupling of $O(3)$-sigma model and the gauge field governed by a Chern-Simons term. In Sec. II, we discuss the $O(3)$-sigma model with the Chern-Simons term subject to the hyperbolic self-dual potential. In Sec. III, we present how the dielectric constant modifies the vortice solutions of the model. Afterwards, we investigate the static vortice solutions in the model and present the respective numerical solutions. Finally, we summarize our results and discussions in Sec. IV.

\section{The minimal gauged $O(3)$ model}

To iniciate this work, we consider the Lagrangean

\begin{align}\label{lagrangian}
\mathcal{L}=\frac{1}{2}D_{\mu}\Phi\cdot D^{\mu}\Phi+\frac{\kappa}{4}\varepsilon^{\mu\nu\alpha}A_{\mu}F_{\nu\alpha}-\mathcal{U}(\Phi_3),
\end{align}
where we defined
\begin{align}
D_{\mu}\Phi=\partial_{\mu}\Phi+ A_{\mu}\hat{n}_{3}\times\Phi.
\end{align}

In the $O(3)$-sigma model, the scalar field $\Phi$ is mapped in the Minkowski space of two unitary spheres, denoted by $S^{2}$. In other words, $\Phi$ is a three-component vector that satisfies the constraint $\Phi\cdot\Phi=\phi_{1}^{2}+\phi_{2}^{2}+\phi_{3}^{2}=1$. Depending on the specific choice for the potential, the Lagrangean may be invariant under a iso-rotation around a preferred axis, in this case,  $\hat{n}_{3}=(0,0,1)$. The $U(1)$ nature of the model can be easely seen in the following identity:
\begin{align}
D_{\mu}\Phi\cdot D^{\mu}\Phi=\vert(\partial_{\mu}+iA_{\mu})(\phi_{1}+i\phi_{2})\vert^{2}+\partial_{\mu}\phi_{3}\partial^{\mu}\phi_{3}.
\end{align}

\begin{figure}[ht]
\centering
\includegraphics[scale=0.9]{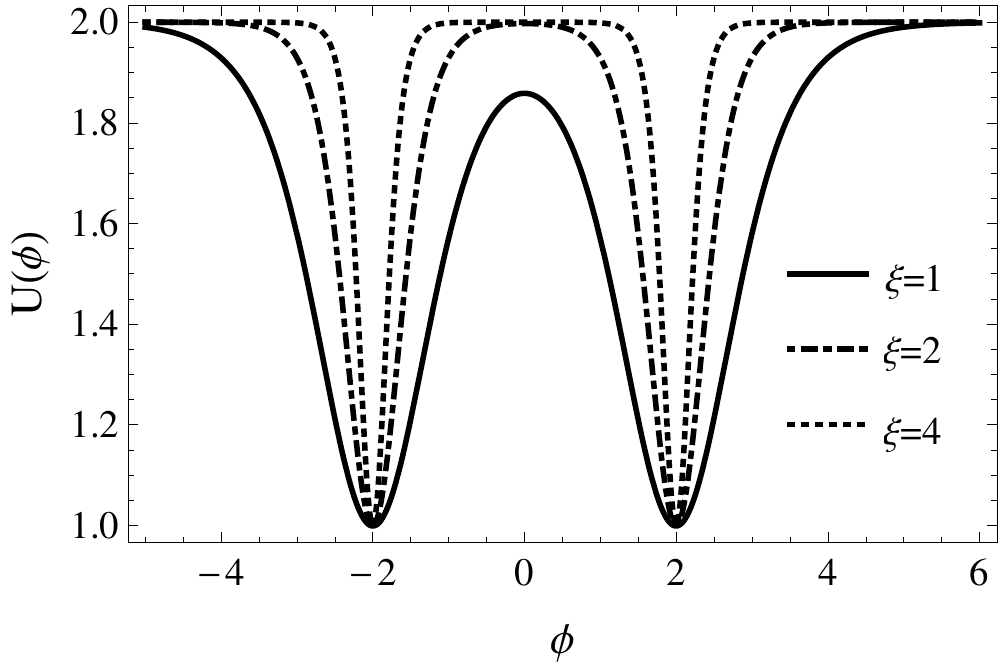}
\caption{Graphical behavior of the potential for several values of the deformation parameter.} \label{fig1}
\end{figure}

To ensure a self-dual system, we consider a potential of the form
\begin{align}\label{potential}
\mathcal{U}(\Phi)=\frac{1}{2\kappa^{2}}\{\tanh^{2}[\xi(\Phi-\Phi_{0})\cdot \hat{n}_{3}]+\tanh^{2}[\xi(\Phi+\Phi_{0})\cdot \hat{n}_{3}]\},
\end{align}
where $\Phi_{0}$ is a constant vector and $\xi$ is a deformation parameter. We can notice that, in this case, the system has two vacuum states, given by $\phi_{0}=\pm\Phi_{0}\cdot\hat{n}_{3}$. See Fig. (\ref{fig1}).

As a result, the $SO(2)$ ($U(1)$) symmetry is not spontaneously broken. In this case, it is also interesting to notice that the gauge field dynamics is regulated only by the Chern-Simons term.

At this point, we should notice that our metric is $\eta_{\mu\nu}=$diag$(+,-,-)$ with $\varepsilon^{012}=1$; $\mu$, $\nu=0,1,2$ and $i$, $j=1,2$. Therefore, the motion equations for the Lagrangean (\ref{lagrangian}) are given by
\begin{align}\label{gauge}
\textbf{J}^{\mu}=-\Phi\times D^{\mu}\Phi,
\end{align}
with $\textbf{J}^{\mu}=-j^{\mu}\cdot\hat{n}_{3}$ and

\begin{align}\label{current_chern_simons}
j^{\mu}=\frac{\kappa}{2}\varepsilon^{\mu\nu\alpha}F_{\nu\alpha}.
\end{align}

For the scalar field, we have
\begin{align}\label{scalar}
D_{\mu}D^{\mu}\Phi=-\frac{\partial\mathcal{U}}{\partial\Phi}.
\end{align}

By replacing  (\ref{gauge}) in (\ref{scalar}), we obtain
\begin{align}\label{euler}
D_{\mu}\textbf{J}^{\mu}=\Phi\times\frac{\partial\mathcal{U}}{\partial\Phi}.
\end{align}

The zeroth component of (\ref{current_chern_simons}) is the well known Gauss law, and thus, the Gauss law implies that the configuration of the field with magnetic flux $\Phi$ carries, essentially, one non-zero charge given by $Q=-\kappa\Phi_{flux}$.

As usual, the energy functional can be obtained by a integration of the component $T_{00}$ of the energy-momentum tensor over all space. In this way, we obtain the functional 
\begin{align}
E=\frac{1}{2}\int d^{2}x \, \bigg[(D_{1}\Phi)^{2}+(D_{2}\Phi)^{2}+\frac{\kappa^{2}F_{12}^{2}}{\phi_{1}^{2}+\phi_{2}^{2}}+2\mathcal{U}\bigg].
\end{align}

Rearranging the energy functional, we obtain
\begin{align}\label{energy}
E=&\frac{1}{2}\int d^{2}x\, \bigg\{ (D_{i}\Phi\pm\varepsilon_{ij}\Phi\times D_{j}\Phi)^{2}+\frac{\kappa^{2}}{\phi_{1}^{2}+\phi_{2}^{2}}\bigg[ F_{12}\mp \sqrt{\frac{2\mathcal{U}(\phi_{1}^{2}+\phi_{2}^{2})}{\kappa^{2}}}\bigg]\bigg\}
\pm 4\pi\int d^{2}x \, \mathcal{Q}_{0}.
\end{align}

Now, we define the topological charge of the model as
\begin{align}
\mathcal{Q}_{\mu}=\frac{1}{8\pi}\varepsilon_{\mu\nu\lambda}\bigg[\Phi\cdot D^{\nu}\Phi\times D^{\lambda}\Phi+F^{\nu\lambda}\sqrt{\frac{2\kappa^{2}\mathcal{U}}{(\phi_{1}^{2}+\phi_{2}^{2})}}\bigg].
\end{align}

Following topological argumentations, we know that the energy (\ref{energy}) has a lower limit \cite{Bogomol'nyi, Atmaja}, i.e., we have
\begin{align}
E\geq 4\pi\int d^{2}x \, \mathcal{Q}_{0}.
\end{align}

At the energy saturation limit, the motion equations of the model reduce to
\begin{align}
D_{i}\Phi\pm\varepsilon_{ij}\Phi\times D_{j}\Phi=0;
\end{align}
\begin{align}
F_{12}\mp \sqrt{\frac{2\mathcal{U}(\phi_{1}^{2}+\phi_{2}^{2})}{\kappa^{2}}}=0.
\end{align}

\subsection{Static vortex solutions}

First, in order to investigate the Bogomol'nyi equation numerically, we choose a spherical rotational symmetry for the variable field \cite{Sales}, that means
\begin{align}\label{ansatz} \nonumber
& \phi_{1}(\rho, \theta)=\sin f(r)\cos N\theta; \\
& \phi_{2}(\rho, \theta)=\sin f(r)\sin N\theta; \\ \nonumber
& \phi_{3}(\rho, \theta)=\cos f(r)
\end{align}
and
\begin{align}
\textbf{A}(\rho, \theta)=-\frac{Na(r)}{\kappa r}\hat{e}_{\theta}.
\end{align}

Considering the self-dual hyperbolic potential, we rewrite the Bogomol'nyi equations as
\begin{align}\label{bogomol'nyi}
f'(r)=\pm N\frac{a+1}{r}\sin f(r);
\end{align}
\begin{align}\label{bogomol'nyi1}
a'(r)=\pm \frac{r}{N}\sqrt{(1-\cos^{2} f(r))\{\tanh^{2}[\xi(\cos f(r)+\phi_{0})]\}+\tanh^{2}[\xi(\cos f(r)-\phi_{0})]\}}.
\end{align}
with $\phi_{0}$ being the vacuum expectation value, $f(r)$ being a arbitrary function and $\rho=r\kappa$ being an adimensional length. $N$ is the parameter responsible for defining the vorticity of the solutions.

By decoupling the previous equations, we find the equation to the variable field
\begin{align}\label{field} 
 f''(r)+\frac{f'(r)}{r}-\frac{f'(r)^{2}}{\tan f(r)}- 
 \sin^{2}f(r)\sqrt{\{\tanh^{2}[\xi(\cos f(r)-\phi_{0})]+\tanh^{2}[\xi(\cos f(r)+\phi_{0})]\}}=0.
\end{align}

It can be easily verified that the Bogomol'nyi equations (\ref{bogomol'nyi}-\ref{bogomol'nyi1}) satisfy the motion equations (\ref{euler}). This is well discussed in the specialized literature, but a more enthusiastic reader can look up at Refs. \cite{Gosh, Cunha, Sales}.

Now, we want to obtain the solutions for equations (\ref{bogomol'nyi}) and (\ref{bogomol'nyi1}). To guarantee that the field is not singular at the origin, we consider that the variable field in the vicinity of the origin has the form: 
\begin{align}
f(0)=n\pi, \, \, \, \, \, n \, \, \in \, \, \mathbb{N},
\end{align}
and then, the regular solutions at the origin require the initial behaviour of the gauge field to be $a(0)=0$. Furthermore, the solutions are symmetric under $f(r)=2\pi$. Therefore, we should have $f(0)=0$ and $f(0)=\pi$ for the topological and non-topological solutions. If we consider the last condition, it is helpful to use the variable change $f(r)=\pi+h(r)$. We also consider the lower sign of the Bogomol'nyi equations and $N$ to be positive. For $h(r)\ll 1$, we assume that the model has solutions of the type
\begin{align}\label{b1}
h(r)=B_{0}r^{N}
\end{align}
and consequently, we obtain
\begin{align}\label{b11}
a(r)=-\frac{B_{0}}{N(N+2)}\sqrt{\tanh[\xi(1+\phi_{0})]^{2}+\tanh[\xi(1-\phi_{0})]^{2}}r^{N+2}+
\mathcal{O}(r^{3N+2}).
\end{align}

On the other hand, if we consider that $f(0)=0$ and $N$ is negative we have, at the vicinity of origin

\begin{align}\label{b2}
f(r)=\bar{B}_{0}r^{-N}
\end{align}
and thus, in this case, the solution for the gauge field is
\begin{align}\label{b22}
a(r)=-\frac{\bar{B}_{0}}{N(2-N)}\sqrt{\tanh[\xi(1+\phi_{0})]^{2}+\tanh[\xi(1-\phi_{0})]^{2}}r^{2-N}+\mathcal{O}(r^{2-3N}).
\end{align}

At infinite, there are two different asymptotic behaviours to the Bogomol'nyi equations. When $f(\infty)=\pi$ and $N$ is positive, $f(r)$ can again be rewritten conveniently as $f(r)=\pi+h(r)$, and then, to obtain the localized energy solutions we should have $a(\infty)=-\eta_{1}$. In this way, we assume that
\begin{align}\label{b3}
h(r)=C_{\infty} r^{N(1-\eta_{1})}.
\end{align}

As a result, we have
\begin{align}\label{b33}
a(r)\simeq -\frac{C_{\infty}}{N[N(1-\eta_{1})+2]}\sqrt{\tanh[\xi(1+\phi_{0})]^{2}+\tanh[\xi(1-\phi_{0})]^{2}}r^{2N(1-\eta_{1})+2}-\eta_{1}.
\end{align} 

To finish, we analyse the boundary condition $f(\infty)=0$ and $a(\infty)=\eta_{2}$ with negative $N$ and we obtain
\begin{align}\label{b4}
f(r)=\bar{C}_{\infty}r^{N(1+\eta_{2})}.
\end{align}
In this way, we obtain the result

\begin{align}\label{b44}
a(r)\simeq -\frac{\bar{C}_{\infty}r^{N(1+\eta_{2})+2}}{N[N(1+\eta_{2})+2]}\sqrt{\frac{1}{2}\{\tanh[\xi(1+\phi_{0})]^{2}+\tanh[\xi(1-\phi_{0})]^{2}\}}+\eta_{2}.
\end{align}
Note that the parameter $\eta_{1}$ ($\eta_{2}$) indicates non-topological (topological) solutions.

\subsection{Numerical results}

\begin{figure}[ht]
\centering
\includegraphics[scale=0.75]{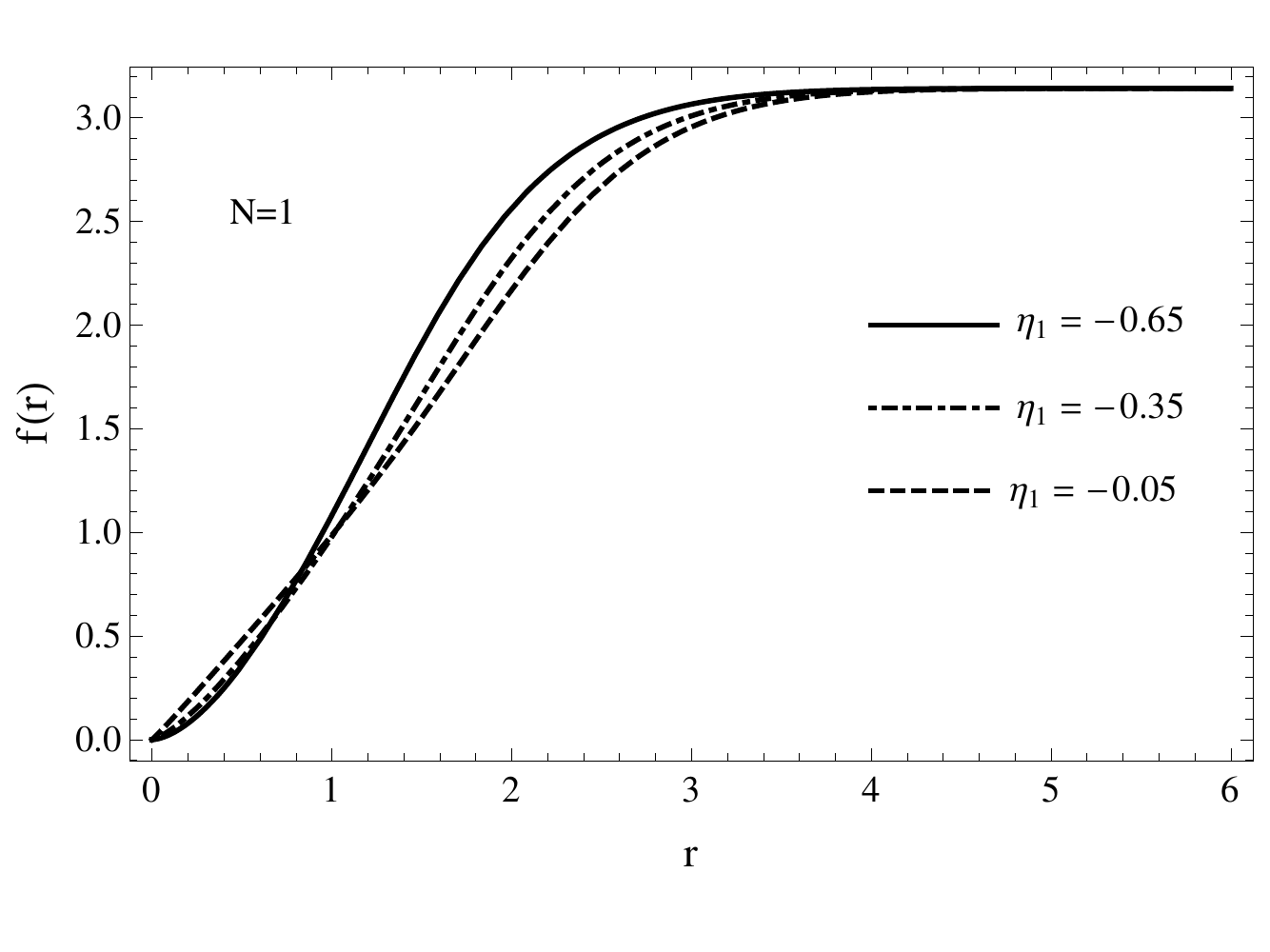}
\vspace{-1cm}
\caption{Numerical solutions of topological vortices for several asymptotic values of the gauge field.}
\label{fig2}
\end{figure}

\begin{figure}[h]
\centering
\includegraphics[scale=0.75]{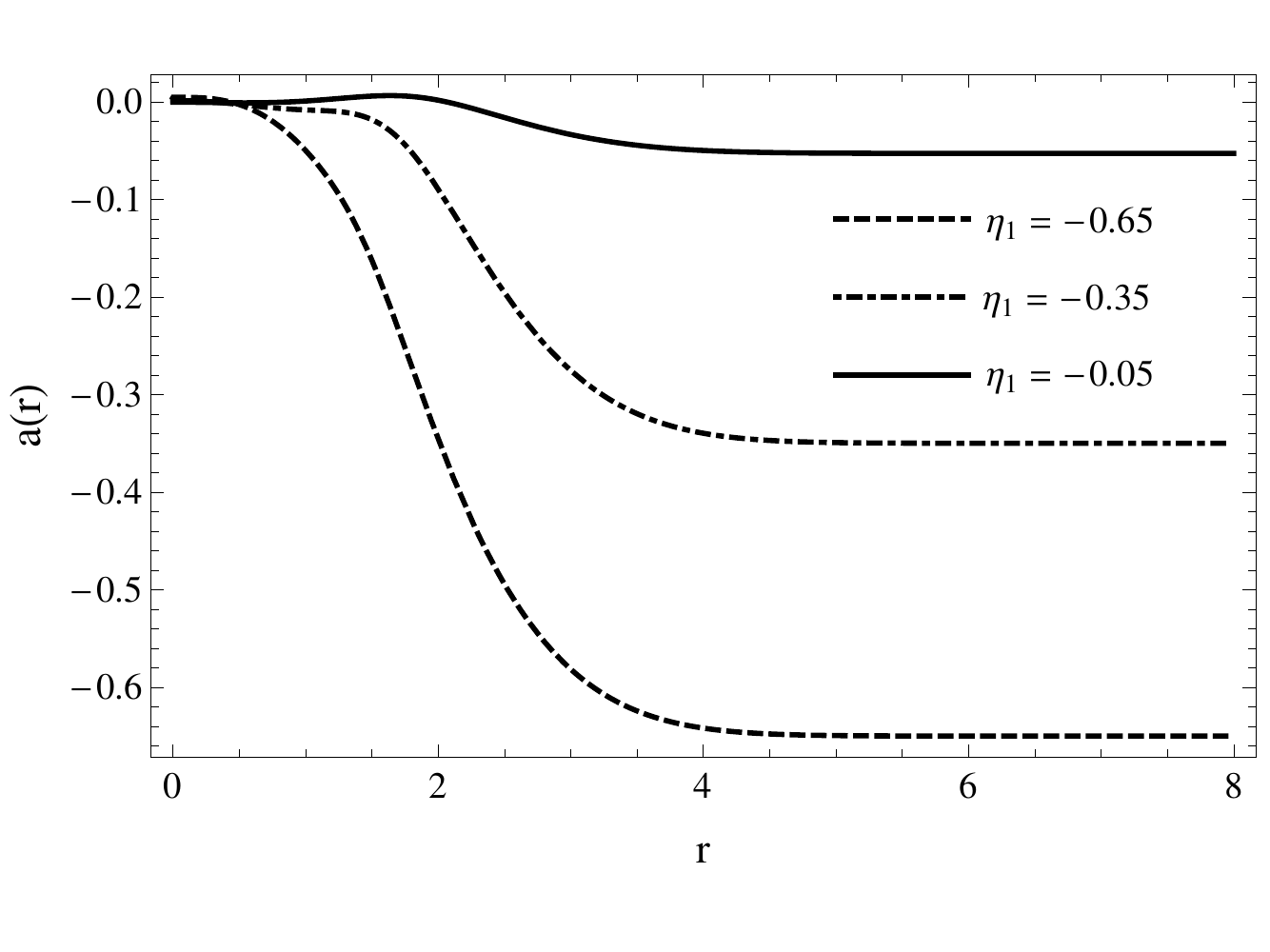}
\vspace{-1cm}
\caption{Behavior of the topological gauge field for different values of the parameter $\eta_{1}$.}
\label{fig3}
\end{figure}

\begin{figure}[ht]
\centering
\includegraphics[scale=0.75]{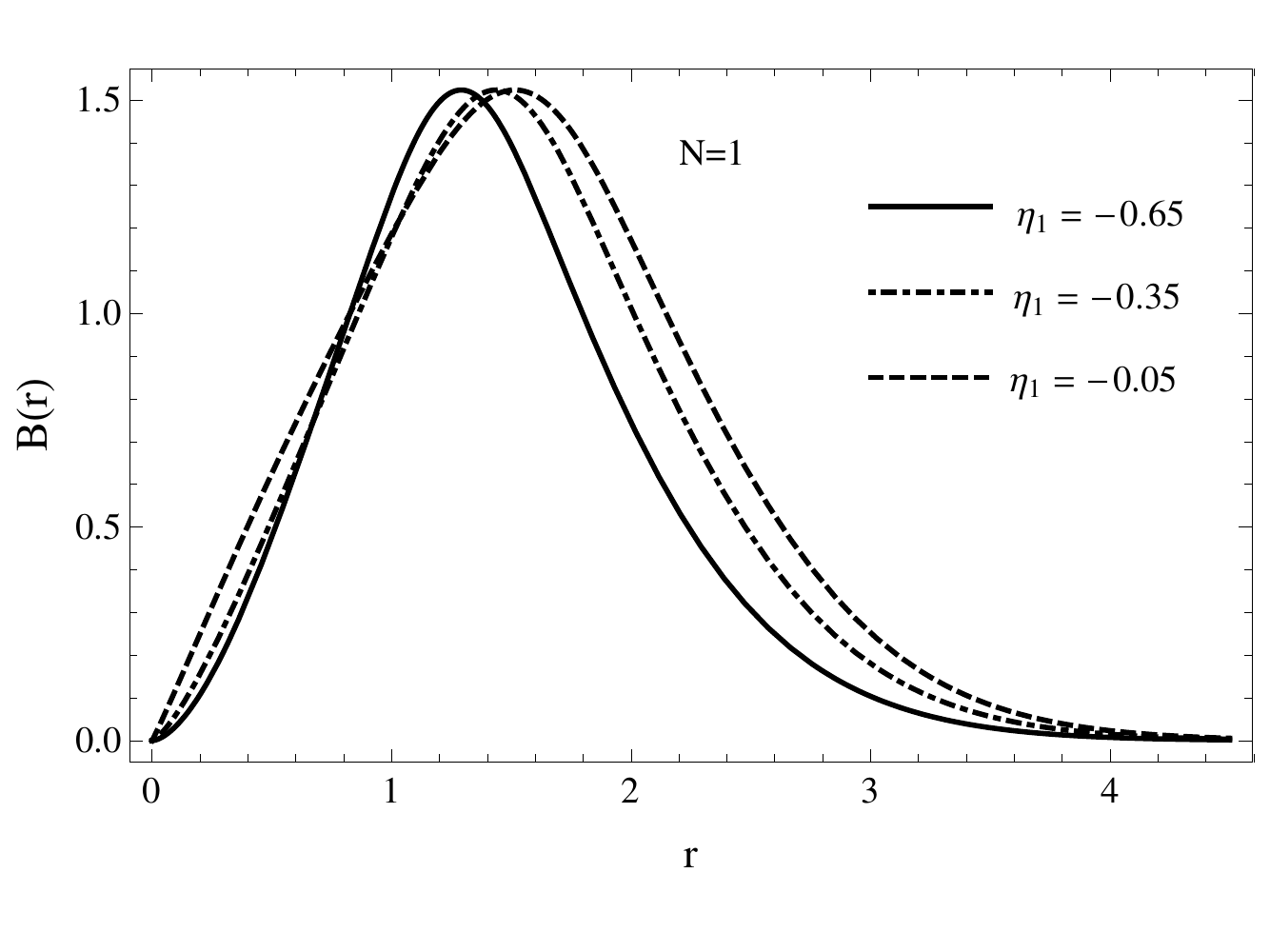}
\vspace{-1cm}
\caption{The magnitude of the magnetic fields $B$ as a function of $r$ for $N=1$.}
\label{fig4}
\end{figure}

From now on, we turn our attention to the numerical analysis of the Bogomol'nyi equations (\ref{bogomol'nyi}-\ref{bogomol'nyi1}). Initially, we investigate numerically the topological solutions with boundary conditions $f(0)=0$ and $f(r\rightarrow\infty)=\pi$ with $N=1$. To obtain the numerical solutions of the coupled Bogomol'nyi equations, we consider the assymptotic behaviours expressed by (\ref{b1}), (\ref{b11}), (\ref{b3}) and (\ref{b33}). In this way, we obtain the numerical results for $f(r)$, $a(r)$ and $B(r)$ as shown, respectively, in the Figs. (\ref{fig2}), (\ref{fig3}) e (\ref{fig4}).

We observe that the variable field and the behaviour of the gauge field are similar to those results presented by a similar model with a non-minimal coupling \cite{Sales}. In this case, since the gauge field is driven exclusively by the Chern-Simons term and the model is subject to the arbitrary hyperbolic potential, the topological solutions are obtained for $N=1$. The behaviour of the magnetic field with $N=1$ is represented in the Fig. (\ref{fig4}).

\begin{figure}[ht]
\centering
\includegraphics[scale=0.75]{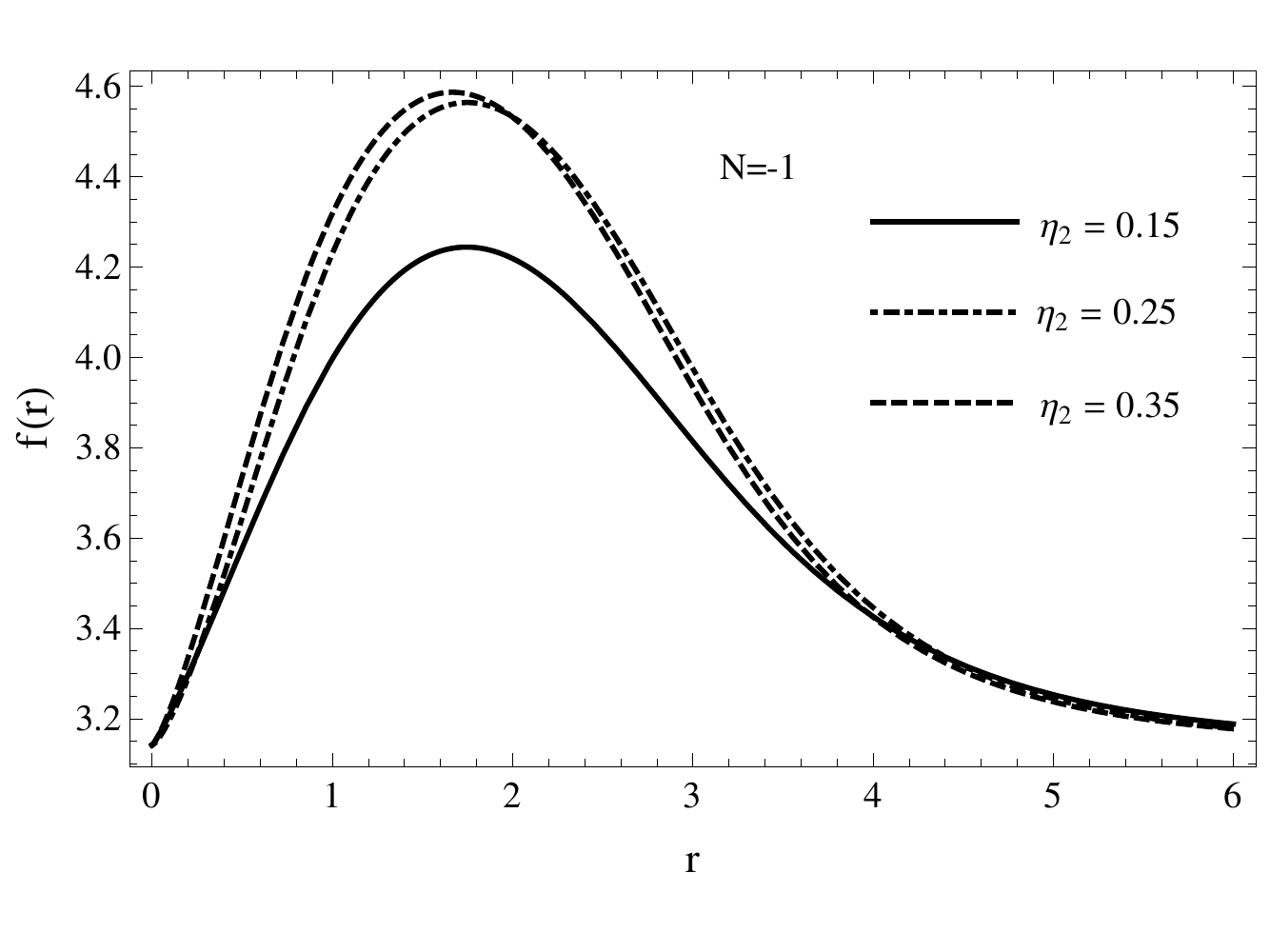}
\vspace{-0.75cm}
\caption{Numerical solutions of nontopological vortices for several asymptotic values of the gauge field.}
\label{fig5}
\end{figure}

\begin{figure}[ht]
\centering
\includegraphics[scale=0.75]{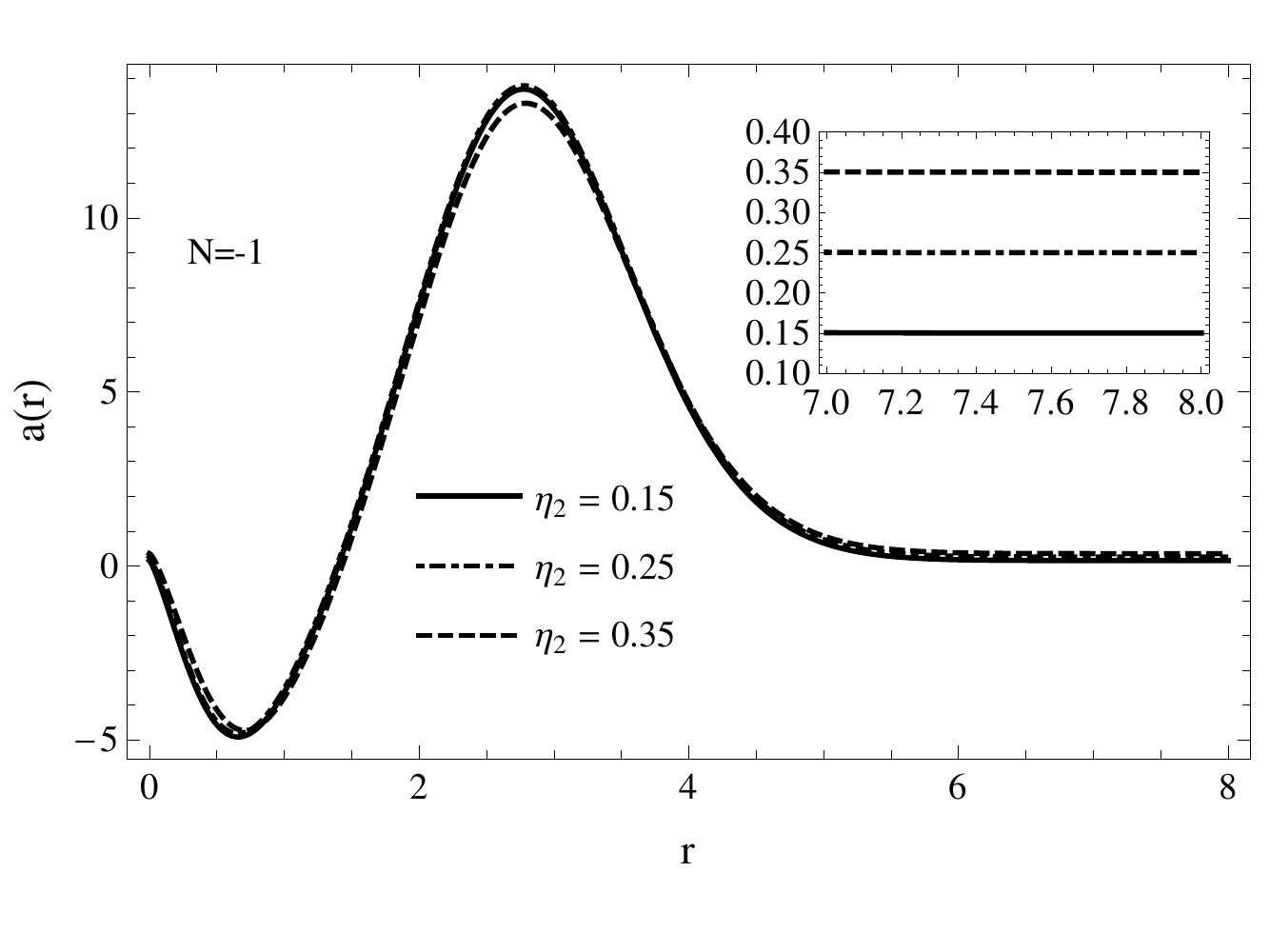}
\vspace{-0.75cm}
\caption{Behaviour of the nontopological gauge field for different values of the parameter $\eta_{2}$.}
\label{fig6}
\end{figure}

\begin{figure}[ht]
\centering
\includegraphics[scale=0.75]{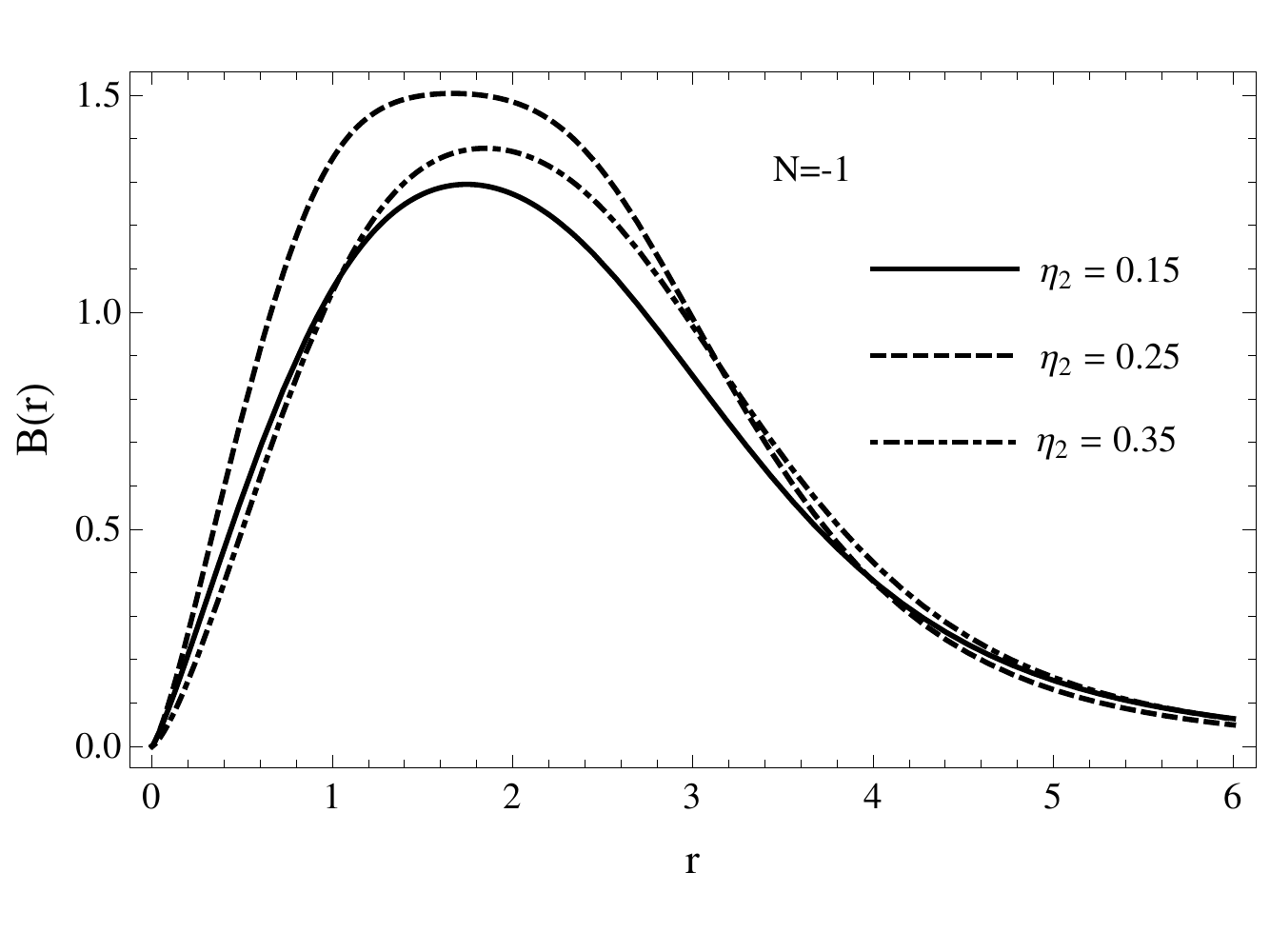}
\vspace{-1cm}
\caption{The magnitude of the magnetic fields $B$ as function of $r$ for $N=-1$.}
\label{fig7}
\end{figure}

For the non-topological solutions, we investigate numerically the existence of solutions with $f(0)=\pi$ and $f(r\rightarrow\infty)=\pi$, $N=-1$ as boundary conditions. With this in mind, in order to find the numerical solutions to the coupled Bogomol'nyi equation, we make use of the assymptotic behaviours expressed by (\ref{b2}), (\ref{b22}), (\ref{b4}) and (\ref{b44}) and thus, obtain the numerical results displayed in Figs. (\ref{fig5}), (\ref{fig6}) and (\ref{fig7}).

We notice that, in the non-topological case, the magnetic field  behaviour resembles the variable field behaviour in the case $N=-1$, with the respective values of the parameter $\eta_{2}$.

\section{The dielectric constant and the compacton-like vortex}
 
In this section, we focus our attention to the discussion of how the dielectric constant modifies the vortice solutions of the variable field. To achieve this aim, we consider the Lagrangean written as
\begin{align}
\mathcal{L}=\frac{1}{2}D_{\mu}\Phi\cdot D^{\mu}\Phi+\frac{\kappa}{4}\omega(\Phi)\varepsilon^{\mu\nu\lambda}A_{\mu}F_{\nu\lambda}-\mathcal{U}(\Phi),
\end{align}
where $\omega$ is a parameter responsible for defining the dielectric constant of the model.

Following arguments similar to those used in the previous section, we once more have the local current given by
\begin{align}
j^{\mu}=\frac{\kappa}{2}\omega(\Phi)\varepsilon^{\mu\nu\lambda}F_{\nu\lambda},
\end{align}
where $\textbf{J}^{\mu}=-j^{\mu}\cdot\hat{n}_{3}$.

The motion equation of the model remains the same
\begin{align}
\textbf{J}^{\mu}= \Phi\times\frac{\partial\mathcal{U}}{\partial\Phi}.
\end{align}

We construct the energy functional and thus, we arrive again at the expression
\begin{align}
E=\frac{1}{2}\int d^{2}x \, \bigg\{(D_{1}\Phi)^{2}+(D_{2}\Phi)^{2}+\frac{\kappa^{2}\omega^{2}F_{12}^{2}}{\phi_{1}^{2}+\phi_{2}^{2}}+2\mathcal{U}\bigg\}.
\end{align}

Rearranging the functional, we obtain the expression
\begin{align}\label{E}
E=\frac{1}{2}\int d^{2}x \, \bigg\{(D_{i}\Phi\pm\varepsilon_{ij}\Phi\times D_{j}\Phi)^{2}+\frac{\kappa^{2}\omega^{2}}{\phi_{1}^{2}+\phi_{2}^{2}}\bigg(F_{12}\mp\sqrt{\frac{2\mathcal{U}(\phi_{1}^{2}+\phi_{2}^{2})}{\kappa^{2}\omega^{2}}}\bigg)^{2}\bigg\}+4\pi\int d^{2}x\, \mathcal{Q}_{0}.
\end{align}

We notice that the topological charge ``acquires" a parameter due to the addition of the dieletric constant. Therefore, we have

\begin{align}
\mathcal{Q}_{\mu}=\frac{1}{8\pi}\varepsilon_{\mu\nu\lambda}\bigg[\Phi\cdot D^{\nu}\Phi\times D^{\lambda}\Phi+F^{\nu\lambda}\sqrt{\frac{2\kappa^{2}\omega^{2}\mathcal{U}}{\phi_{1}^{2}+\phi_{2}^{2}}}\bigg].
\end{align}

Once again we use the argumentation of the previous section and write the Bogomol'nyi equations as

\begin{align}
D_{i}\Phi\pm\varepsilon_{ij}\Phi\times D_{j}\Phi=0;
\end{align}
\begin{align}
F_{12}\mp\sqrt{\frac{2\mathcal{U}(\phi_{1}^{2}+\phi_{2}^{2})}{\kappa^{2}\omega^{2}}}=0.
\end{align}

Considering the ansatz (\ref{ansatz}), we obtain the coupled expression

\begin{align}
f'(r)=\pm N\frac{a+1}{r}\sin f(r)
\end{align}
and
\begin{align}
a'(r)=\pm \frac{r}{\omega N}\sqrt{(1-\cos^{2}f(r))(\tanh^{2}[\xi(\cos f(r)-1)]+\tanh^{2}[\xi(\cos f(r)+1)]}.
\end{align}

Decoupling the previous equations, we obtain

\begin{align}\label{field1} 
f''(r)+\frac{f'(r)}{r}-\frac{f'(r)^{2}}{\tan f(r)}-\frac{1}{\omega}\sin ^{2}f(r)\sqrt{\{\tanh^{2}[\xi(\cos f(r)-\phi_{0})]+\tanh^{2}[\xi(\cos f(r)+\phi_{0})]\}}=0.
\end{align}

The equation (\ref{field1}) is analogous to the equation (\ref{field}) when the parameter $\omega=1$. From now on, we focus on understanding how the dielectric constant modifies the variable field solutions. In order to do this, we iniciate the numerical study of the equation (\ref{field1}) using the above-mentioned boundaries.

\subsection{Numerical study of dielectric constant}

First, in order to investigate the influence of the dielectric constant on the solutions of the variable field $f(r)$, we use a numerical approach. We consider the numerical solution presented in Fig. (\ref{fig2}) with $a(\infty)=-0.35$. Then, we vary the parameter $\omega$. As a result, we observe that the kink-like solutions acquire a characteristic of a compacton-like solution by simply varying numerically the parameter $\omega$ (dielectric constant) of the model. In other words, as the gauge field or the Chern-Simons term contributions decrease, the kink-like solutions become compacton-like solutions, as shown in Fig. ({\ref{fig8}).

\begin{figure}[ht]
\centering
\includegraphics[scale=0.75]{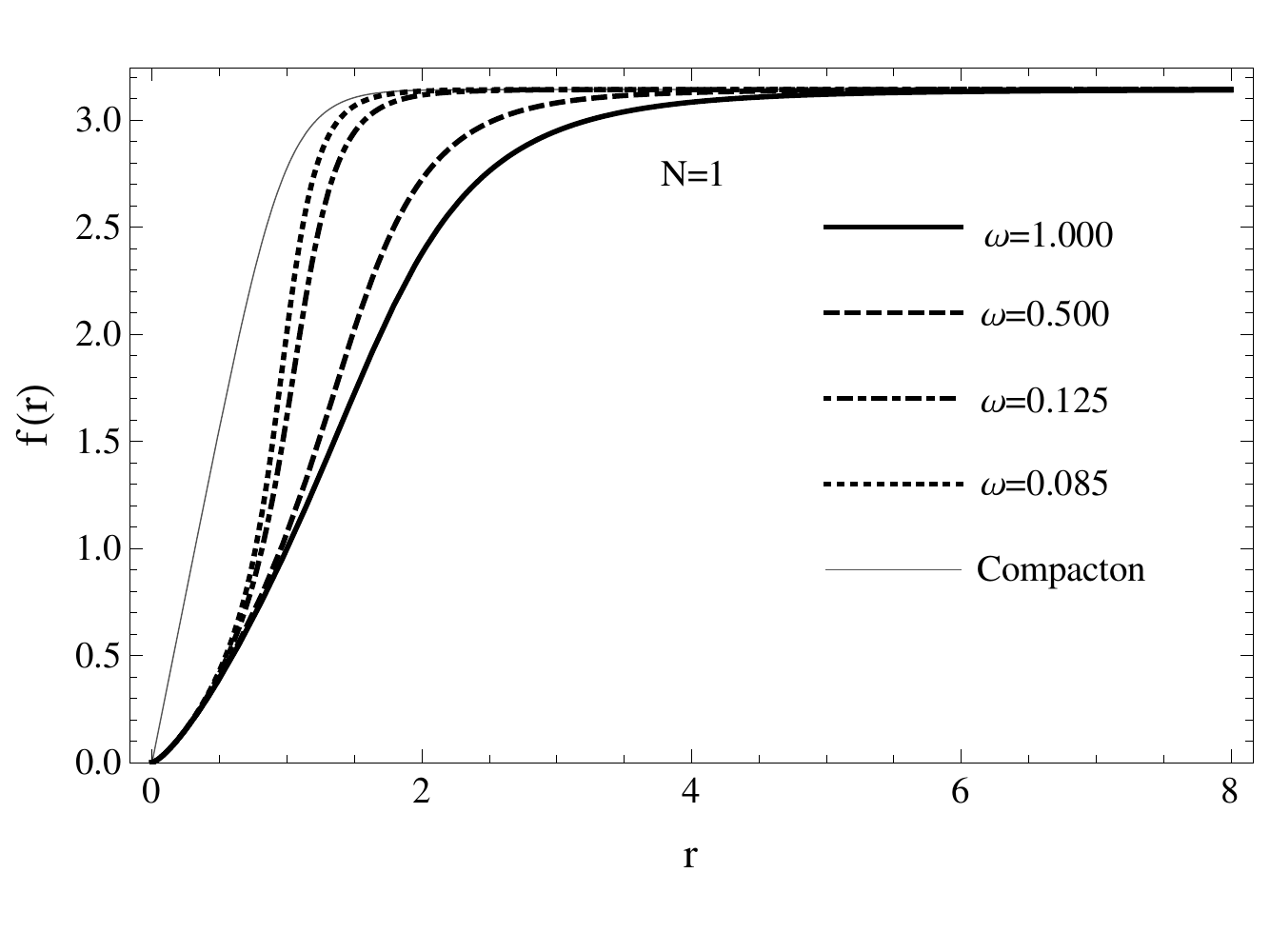}
\vspace{-1cm}
\caption{From the kink-like topological solutions to compacton-like solutions.}
\label{fig8}
\end{figure}

Interesting results also emerge when we vary numerically $\omega$ in the non-topological solutions presented in Fig. (\ref{fig5}). For this result, we consider $\eta_2=0.25$ and vary the parameter $\omega$. The outcome of these numerical results are shown in Fig. (\ref{fig9}). In this case, we notice that when we increase the gauge field contribution, the solitonic solutions approach the kink-like solutions for the variable field.

\begin{figure}[ht]
\centering
\includegraphics[scale=0.75]{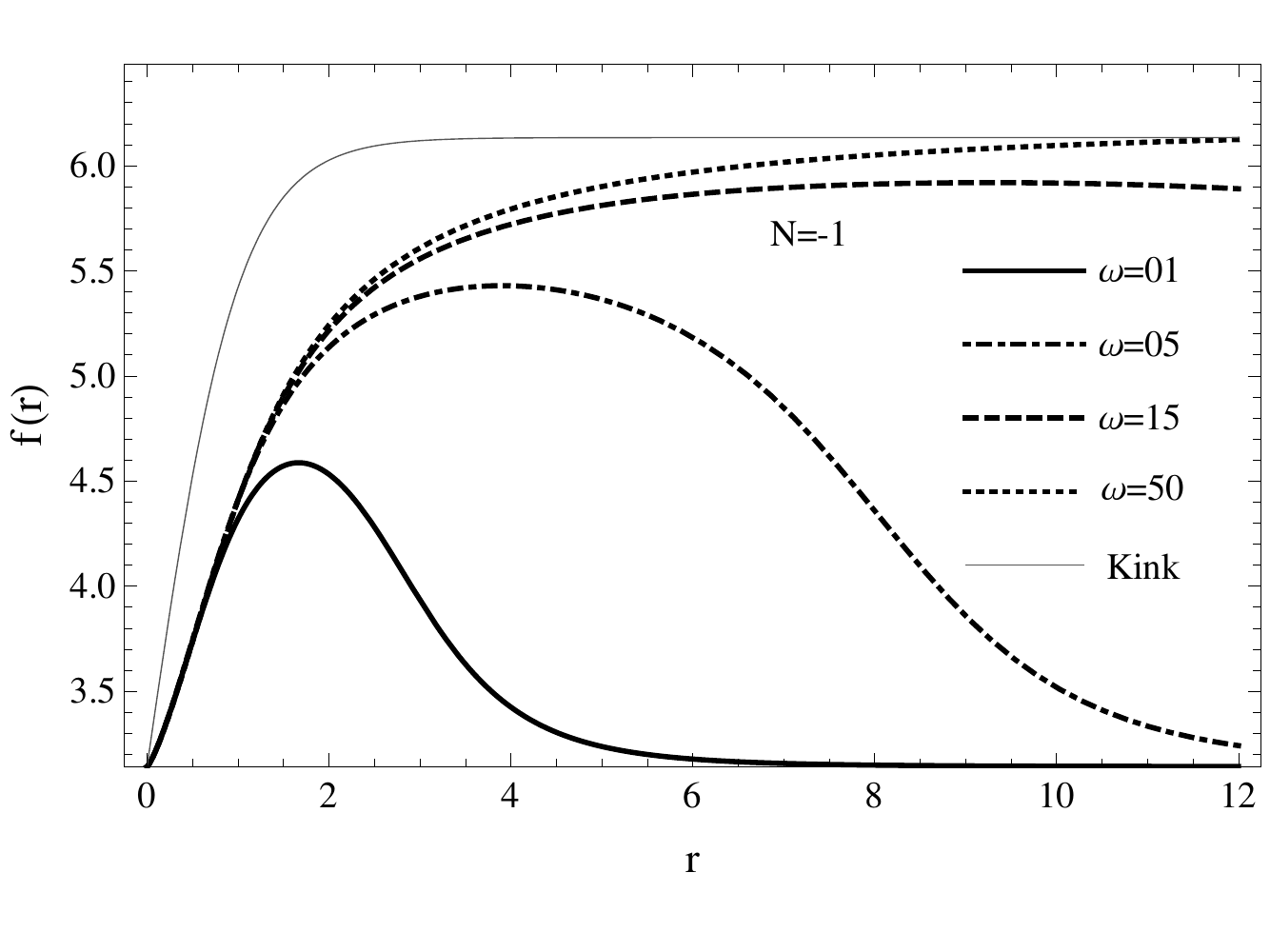}
\vspace{-1cm}
\caption{Nontopological kink-like solutions for the variable field.}
\label{fig9}
\end{figure}

\section{Concluding remarks}

In this work, we investigate the vortices solutions of the $O(3)$-sigma model with the gauge field coupled minimally with the Chern-Simons term. As a result, we note that due to the ansatz and the constraint of the $O(3)$ model, for the boundary conditions $f(0)=0$ and $f(\infty)=\pi$, we have the so-called non-topological solutions, since the topological charge is given by a non-integer parameter. On the other hand, for the boundary conditions $f(0)=\pi$ and $f(\infty)=\pi$ we have topological solutions, which have a topological charge described by an integer. As a consequence, the topological vortices of the model have a quantized energy given by $\mathcal{E}= 4\pi\vert N\vert$, the charge being $Q=-\kappa\Phi_{flux}$ and the magnetic flux given by $\Phi_{flux}=2\pi N\eta_1$. Particularly, we note that although the energy is quantized, the flux of the model is not quantized. We also observe that for a fixed $N$ there exists a family of solutions for different values of $\eta_{1}$, which implies that there are infinitely many degenerate solutions in the model. The so-called non-topological solutions are characterized by an energy $\mathcal{E}=4\pi N\eta_2$, by a flux $\Phi_{flux}=2\pi N\eta_2$ and  by a charge $Q=-\kappa\Phi_{flux}$. Finally, after modifying the model by the introduction of a dielectric constant, we note that through the their variation, we can take along  kink-like solutions (topological) to compacton-like solutions by a numerical variation of the dielectric constant of the model. In other words, as the contribution of the gauge field decreases, the contribution of the Chern-Simons terms also decreases, so we have that solutions kink-like become solutions compacton-like, as shown in Fig. (\ref{fig8}). In the same way, we observe that when we increase the contribution of the gauge field in the non-topological solutions, the solitonic solutions approach the kink-like solutions for the variable field and, therefore, the non-topological solutions tend to topological solutions with $\vert Q_{top}\vert=N$.

\section*{Acknowledgments}

\hspace{0.5cm}The authors would like to thank the Funda\c{c}\~{a}o Cearense de apoio ao Desenvolvimento Cient\'{\i}fico e Tecnol\'{o}gico (FUNCAP) the Coordena\c{c}\~ao de Aperfei\c{c}oamento de Pessoal de N\'ivel Superior (CAPES), and the Conselho Nacional de Desenvolvimento Cient\'{\i}fico e Tecnol\'{o}gico (CNPq) for financial support.

\end{document}